\documentclass[12pt]{article}
\begin{document}
\title{String Theory: An Evaluation}
\author{Peter Woit \\
Department of Mathematics, Columbia University\\
woit@math.columbia.edu}
\maketitle

For nearly seventeen years now most speculative and mathematical work in
particle theory has centered around the idea of replacing quantum field
theory with something that used to be known as \lq\lq Superstring Theory", 
but
now goes under the name \lq\lq M-Theory". We've been told that 
\lq\lq string 
theory is
a part of twenty-first-century physics that fell by chance into the
twentieth century", so this year the time has perhaps finally come to
begin to evaluate the success or failure of this new way of thinking about
particle physics.  This article will attempt to do so from the perspective
of a quantum field theorist now working in the mathematical community.

The theory has been spectacularly successful on one front, that of public
relations.  Best-selling books and web sites are devoted to explaining the
subject to the widest possible audience. The NSF is funding a series of
NOVA programs on string theory, and the ITP at Santa Barbara is organizing
a conference to train high school teachers in string theory so that they
can teach it to their students.  The newspaper of record informs us that
\lq\lq Physicists Finally Find a Way to Test Superstring Theory" (NYT 
4/4/00).

The strongest scientific argument in favor of string theory is that it
appears to contain a theory of gravity embedded within it.  It is not
often mentioned that this is not yet a consistent quantum theory of
gravity. All that exists at the moment is a divergent series that is
conjectured to be an asymptotic perturbation series for some as yet
undefined non-perturbative string theory (the terms in the series are
conjectured to be finite, unlike the situation in the standard
quantization of general relativity).  String theorists actually consider
the divergence of this series to be a virtue, since otherwise they would
have an infinity (one for each compactification of six dimensions) of
consistent theories of gravity on their hands, with no principle for
choosing amongst them.

String theory has lead to many striking new mathematical results. The
concept of \lq\lq mirror symmetry" has been very fruitful in algebraic 
geometry,
and conformal field theory has opened up a new, fascinating and very deep
area of mathematics.  Unfortunately the mathematically interesting parts
of string theory have been pretty much orthogonal to those parts that
attempt to connect with the real world.
 
The experimental situation is best described with Pauli's phrase 
\lq\lq it's not
even wrong".  No one has managed to extract any sort of experimental
prediction out of the theory other than that the cosmological constant
should probably be at least 55 orders of magnitude larger than
experimental bounds.  String theory not only makes no predictions about
physical phenomena at experimentally accessible energies, it makes no
predictions whatsoever. Even if someone were to figure out tomorrow how to
build an accelerator capable of reaching Planck-scale energies, string
theorists would be able to do no better than give qualitative guesses
about what such a machine might see.  This situation leads one to question
whether string theory really is a scientific theory at all.  At the moment
it's a theory that cannot be falsified by any conceivable experimental
result. It's not even clear that there is any possible theoretical
development that would falsify the theory.

String theorists often attempt to make an aesthetic argument, a claim that
the theory is strikingly \lq\lq elegant" or \lq\lq beautiful".  Since 
there is no
well-defined theory, it's hard to know what to make of these claims, and
one is reminded of another quote from Pauli.  Annoyed by Heisenberg's
claims that modulo some details he had a wonderful unified theory (he
didn't), Pauli sent his friends a postcard containing a blank rectangle
and the text \lq\lq This is to show the world I can paint like Titian. 
Only
technical details are missing." Since no one knows what \lq\lq M-theory" 
is, its
beauty is that of Pauli's painting.  Even if a consistent M-theory can be
found, it may very well be a theory of great complexity and ugliness.

From a mathematician's point of view, the idea that M-theory will replace
the Standard Model with something aesthetically more impressive is rather
suspicious.  Two of the most important concepts of the Standard Model are
that of a gauge field and that of the Dirac operator.  Gauge fields are
identical with connections, perhaps the most important objects in the
modern formulation of geometry.  Thinking seriously about the infinite
dimensional space of all connections has been a very fruitful idea that
mathematicians have picked up from physicists.  The importance of the
Dirac operator is well known to physicists, what is less well known is
that it is of similar importance in mathematics where it plays the role of
\lq\lq fundamental class" in K-theory.  This is reflected in the central role
the Dirac operator plays in the Atiyah-Singer index theorem, one of the
great achievements of twentieth century mathematics.

To the extent that the conceptual structure of string theory is
understood, the Dirac operator and gauge fields are not fundamental, but
are artifacts of the low energy limit. The Standard Model is dramatically
more \lq\lq elegant" and \lq\lq beautiful" than string theory in that its 
crucial
concepts are among the deepest and most powerful in modern mathematics.
String theorists are asking mathematicians to believe in the existence of
some wonderful new mathematics completely unknown to them involving
concepts deeper than that of a connection or a Dirac operator.  This may
be the case, and one must take this argument seriously when it is made by
a Fields medalist, but without experimental evidence or a serious proposal
for what M-theory is, the argument is unconvincing.

Given the lack of experimental or aesthetic motivation, why do so many
particle theorists work on string theory?  Sheldon Glashow describes
string theory as \lq\lq the only game in town", but this begs the 
question.  
Why is it the only game in town?

During much of the twentieth century there were times when theoretical
particle physics was conducted quite successfully in a somewhat faddish
manner; there was often only one game in town.  Experimentalists regularly
discovered new unexpected phenomena, each time leading to a flurry of
theoretical activity and sometimes to Nobel prizes for those quickest to
correctly understand the significance of the new data.  Since the
discovery of the $J/\Psi$ in November 1974, there have been no solid
experimental results that disagree with the Standard Model (except perhaps 
recent indications of neutrino masses).  It is likely
that this situation will continue at least until 2006 when experiments at
the LHC at CERN are scheduled to begin.  To a large extent particle theory
research has continued to be conducted in a faddish manner for the past
quarter century, but now with little success.

Graduate students, post-docs and untenured junior faculty interested in
physics beyond the Standard Model are under tremendous pressures in a
brutal job market to work on the latest fad in string theory, especially
if they are interested in speculative and mathematical research. For them,
the idea of starting to work on an untested new idea that may very well
fail looks a lot like a quick route to professional suicide.  Many physics
researchers do not believe in string theory but work on it anyway. They
are often intimidated intellectually by the fact that some leading string
theorists are undeniably geniuses, and professionally by the desire to
have a job, get grants, go to conferences and generally have an
intellectual community in which to participate.

What can be done? Even granting that string theory is an idea that
deserves to be pursued, how can theorists be encouraged to try and find
more promising alternatives?  Here are some modest proposals, aimed at
encouraging researchers to strike out in new directions:

1. Until such time as a testable prediction (or even a consistent
compelling definition) emerges from string theory, theorists should
publicly acknowledge the problems theoretical particle physics is facing,
and should cease and desist from activities designed to sell string theory
to impressionable youths, popular science reporters and funding agencies.

2. Senior theorists doing string theory should seriously reevaluate their
research programs, consider working on less popular ideas and encourage
their graduate students and post-docs to do the same.

3. Instead of trying to hire post-docs and junior faculty working on the
latest string theory fad, theory groups should try and identify young
researchers who are working on original ideas and hire them to long enough
term positions that they have a chance of making some progress.

4. Funding agencies should stop supporting theorists who propose to
continue working on the same ideas as everyone.  They should also question
whether it is a good idea to fund a large number of conferences and
workshops on the latest string theory fad.  Research funds should be
targeted at providing incentives for people to try something new and
ambitious, even if it may take many years of work with a sizable risk of
ending up with nothing.

Particle theorists should be exploring a wide range of alternatives to
string theory, and looking for inspiration wherever it can potentially be
found.  The common centrality of gauge fields and the Dirac operator in
the Standard Model and in mathematics is perhaps a clue that any
fundamental physical model should directly incorporate them.  Another
powerful and unifying idea shared by physics and mathematics is that of a
group representation. Some of the most beautiful mathematics to emerge
from string theory involves the study of (projective) representations of
the group of conformal transformations and of one-dimensional gauge groups
(\lq\lq loop groups").  This work is essentially identical with the study 
of two
dimensional quantum field theory.  The analogous questions in four
dimensions are terra incognita, and one of many potentially promising
areas particle theorists could look to for inspiration.

During the 1960's and early 1970's, quantum field theory appeared to be
doomed and string theory played a leading role as a theory of the strong
interactions. Could it be that just as string theory was wrong then, it is
wrong now, and in much the same way:  perhaps the correct quantum theory
of gravity is some form of asymptotically free gauge theory?  As long as
the best young minds of the field are encouraged to ignore quantum field
theory and pursue the so far fruitless search for M-theory, we may never
know. 

\end{document}